\documentclass[aps,prb,twocolumn,showpacs,floats,epsfig]{revtex4}
\usepackage{epsfig,psfrag,subfigure,amsopn}
\usepackage{graphicx,psfrag,subfigure}
\usepackage{amsmath,amssymb,amsbsy}
\usepackage{dsfont}
\usepackage{subfigure}
\usepackage[dvips]{color}
\usepackage{bbold}
\newcommand\bea{\begin{eqnarray}}
\newcommand\eea{\end{eqnarray}}
\newcommand\beq{\begin{equation}}
\newcommand\eeq{\end{equation}}

\newcommand{\noi}{\noindent}
\newcommand{\bib}{\bibitem}

\def\non{\nonumber}

\def\al{\alpha}

\def\de{\delta}
\def\ep{\epsilon}
\def\ga{\gamma}
\def\Ga{\Gamma}
\def\lam{\lambda}

\def\si{\sigma}
\def\vp{v_\parallel}
\def\vz{v_z}
\def\pa{\partial}

\def\dg{\dagger}
\def\la{\langle}
\def\ra{\rangle}

\def\th{\theta}

\begin{document}

\title{Edge States of a Three-dimensional Topological Insulator}

\author{Oindrila Deb$^1$, Abhiram Soori$^2$, and Diptiman Sen$^1$}
\affiliation{$^1$Centre for High Energy Physics, Indian Institute of Science,
Bangalore 560 012, India \\
$^2$Max Planck Institute for the Physics of Complex Systems, N\"othnitzer 
Str. 38, 01187 Dresden, Germany }

\begin{abstract}
We use the bulk Hamiltonian for a three-dimensional topological insulator 
such as $\rm Bi_2 Se_3$ to study the states which appear on its various 
surfaces and along the edge between two surfaces. We use both analytical
methods based on the surface Hamiltonians (which are derived from the bulk 
Hamiltonian) and numerical methods based on a lattice discretization of the 
bulk Hamiltonian. We find that the application of a potential barrier 
along an edge 
can give rise to states localized at that edge. These states have an unusual 
energy-momentum dispersion which can be controlled by applying a potential 
along the edge; in particular, the velocity of these states can be tuned to 
zero. The scattering and conductance across the edge is studied as a function 
of the edge potential. We show that a magnetic field in a particular direction
can also give rise to zero energy states on certain edges. We point out 
possible experimental ways of looking for the various edge states.
\end{abstract}

\maketitle

\section{Introduction}~\label{sec:intro}
The last few years have witnessed extensive studies, both 
theoretical~\cite{zhang1,kane1,kane2,teo,qi} and 
experimental~\cite{hasan1,exp2,exp1,hasan2} of a class of materials called 
topological insulators (TI). These are materials which, ideally, have only 
gapped states in the bulk and gapless states on the boundaries which are 
protected by time-reversal symmetry~\cite{tirev1,tirev2}. A two-dimensional 
TI hosts one-dimensional gapless edge states while a three-dimensional 
TI hosts gapless states on its two-dimensional surfaces.
Materials such as $\rm Bi_2 Te_3$ and $\rm Bi_2 Se_3$ are known to have
surfaces which host a single Dirac cone near the $\Gamma$ point of the 
surface Brillouin zone~\cite{hasan1,exp2,hasan2}. Many interesting features 
of the surface states have been studied~\cite{kane4,tirev1,tirev2,fu,been1,
tanaka1,tanaka2,mondal1,hasan2,burkov,garate,zyuzin,das1}. Some of these 
studies 
examined interfaces involving a TI and proximate magnetic or superconducting 
materials~\cite{kane4,been1,tanaka1,tanaka2,burkov,garate,guigou,das1,rein}.
Junctions of different surfaces of TIs (which are sometimes separated by a 
geometrical step or a magnetic domain wall)~\cite{taka,deb,wickles,biswas,
alos,sitte,zhang3,apalkov,habe}, polyhedral surfaces~\cite{ruegg}, and 
junctions of surfaces of a TI with normal metals or magnetic 
materials~\cite{modak} or superconductors~\cite{soori,nuss} have also been 
studied. Effects of finite 
sizes~\cite{shen,linder,egger,kundu,shenoy,pertsova,neupane} and different 
orientations~\cite{zhang3,habe,silvestrov,takane,barreto,rao,brey} on 
the surface states have been studied. Zero energy states at the edges of a thin
strip have been studied in Ref.~\onlinecite{paananen1} and it has been shown 
that these remain robust in the presence of Zeeman coupling to a magnetic 
field. Zero energy surface states produced by a Zeeman field have been studied
in Ref.~\onlinecite{paananen2}. Finally, transport around different surfaces 
of a TI in the presence of a magnetic field and the Aharonov-Bohm effect has 
been experimentally studied in Ref.~\onlinecite{peng}. 

The main motivation of our work is to start from the bulk Hamiltonian of a 
three-dimensional TI like $\rm Bi_2 Se_3$ and study the states which appear 
at the edge between two cleaved surfaces of the system and the effects of a 
magnetic field and a barrier potential on such edge states. Although the edge
states can be studied more easily starting from the surface Hamiltonians 
and using some boundary conditions (as we will discuss below), the advantages
of obtaining the edge states from the bulk Hamiltonian are the following. 

\noi (i) It would confirm that the edge states are not an artifact of the 
boundary conditions on the surface Hamiltonians.

\noi (ii) It gives us a better understanding of the physical conditions 
necessary for the appearance of edge states. 

\noi Since analytical calculations starting from the bulk 
Hamiltonian are not feasible for this problem, we will introduce a lattice 
discretization of the Hamiltonian and carry out calculations numerically.

The plan of the paper is as follows. In Sec.~\ref{sec:continuum} we review 
how the states on the different surfaces of the system can be obtained from 
a continuum version of the bulk Hamiltonian which involves four bands 
(including spin)~\cite{zhang3}. We will be interested in 
a system which is infinitely long in, say, the $y$ direction (so that the 
momentum $k_y$ is a good quantum number) and has a cross-section which is a 
square lying in the $x-z$ plane~\cite{peng}. Hence 
the system has four surfaces which are separated pairwise by four infinitely 
long edges. We show that although the Hamiltonians governing the states 
on different surfaces have quite different forms (as a result of which the 
dispersion and spin-momentum locking look quite different), the wave 
functions on different surfaces can be related by a unitary transformation 
which conserves the current perpendicular to the edges. We then consider 
the scattering of the surface states from the edges. In the absence of a 
potential applied along an edge, we discover that there is no reflection at 
that edge, but if a potential is applied, there is reflection. 
We calculate the conductance across an edge as 
a function of the potential at the edge. Next, we show that states
localized along an edge can appear in two situations. If a potential is 
applied along an edge, edge states appear on that edge. Their wave functions 
exponentially decay away from that edge into the adjoining
surfaces. Rather unusually, the dispersion of these edge states
is found to be of the form $E = v |k_y|$, where $k_y$ is the momentum along
the edge; the spin structure of these states is also different from that of
the surface states. Both the sign and magnitude of $v$ can be controlled by
varying the strength of the potential at the edge. In particular, $v$ can 
be tuned to zero
thereby giving rise to edge states with zero group velocity. We also discover 
that the presence of a magnetic field in the $x$ direction can produce states 
on two out of the four edges for certain ranges of values of $k_y$. These 
states have zero energy and their spin points along the $z$ direction.

To understand the various edge states better, we then study them numerically
as follows. In Sec.~\ref{sec:lattice} we introduce a lattice 
discretization of the bulk Hamiltonian in the $x-z$ 
plane, maintaining the continuum approximation in the $y$ direction. In
Sec.~\ref{sec:numerics} we present our numerical results for the various 
states of the system. These are of two types: bulk states which are gapped and
surface states which are gapless (in the absence of a magnetic field) with a 
Dirac spectrum. If a potential is applied along an edge or a magnetic field is 
applied in the $x$ direction, we discover that a third type of states appear; 
these are the edge states whose wave functions are localized along an edge and
exponentially decay away into both the adjoining surfaces and the bulk. The 
dispersion and spin structure of these edge states agree qualitatively with 
that found in Sec.~\ref{sec:continuum}. In Sec.~\ref{sec:summary}, we 
summarize our work and discuss some possible experiments to test our results.

\begin{figure}[t]
\epsfig{figure=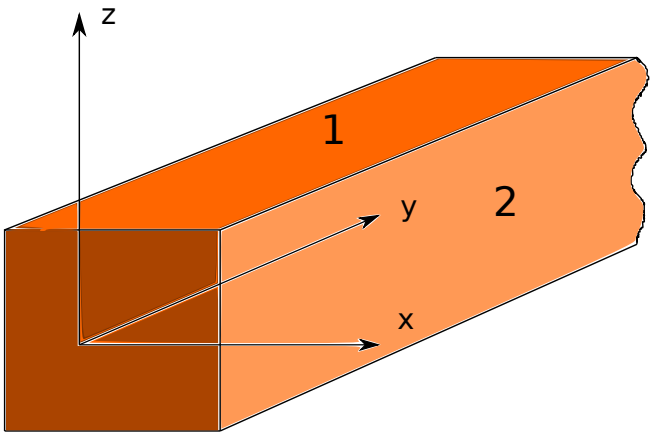,width=8.6cm}
\caption{(Color online) Schematic picture of a system with four surfaces with 
finite widths and infinite length. The top surface (1), the side surface (2), 
and the junction between them will be studied in this paper.} 
\label{fig01} \end{figure}

\section{Continuum model}~\label{sec:continuum}
In this section, we review how the surface states of a TI like $\rm 
Bi_2 Se_3$ arise from a four-component continuum description of the 
bulk~\cite{zhang3}. We consider a Hamiltonian in the bulk of the form 
\beq H_b ~=~ - m \tau^z ~-~ i \vz \tau^y \pa_z ~+~ i \vp \tau^x ~
(\si^y \pa_x ~-~ \si^x \pa_y), \label{ham1} \eeq
where the constants $m$, $\vz$ and $\vp$ are all positive. Here $\si^a$ 
denote the Pauli spin matrices, while $\tau^a$ denotes pseudospin, with 
$\tau^z = \pm 1$ denoting $Bi$ and $Se$ respectively. (We will set $\hbar =1$ 
unless mentioned otherwise). In a basis in which $\si^y$ and $\tau^y$ are 
imaginary and the other four matrices are real, the above Hamiltonian is 
time-reversal symmetric where the time-reversal transformation complex 
conjugates all numbers and transforms wave functions as $\psi \to \si^y 
\psi^*$.

For a translationally invariant system in which momentum is a good quantum 
number, the dispersion following from Eq.~\eqref{ham1} is given by
\beq E_{k_x,k_y,k_z} ~=~ \pm ~\sqrt{\vp^2 (k_x^2 + k_y^2) ~+~ \vz^2
k_z^2 ~+~ m^2}, \label{disp1} \eeq
which has a gap of $2m$ at zero momentum.

\subsection{Surface states}

We now examine how surface states arise from Eq.~\eqref{ham1}. 
Ref.~\onlinecite{zhang3} has discussed this for plane surfaces with arbitrary
orientations. A schematic picture of a system with a square cross-section
is shown in Fig.~\ref{fig01}; the surfaces are infinitely long in the 
$y$ direction and have a finite width in the other two directions. 
We will only consider two of the surfaces in this section, the
top surface and a side surface, which we label as 1 and 2 respectively,
as shown in Fig.~\ref{fig01}. 

We take the top surface to lie at $z=0$ such that the system fills up the
region with $z<0$ and there is vacuum in the region with $z>0$. To define the 
problem completely, we have to specify the boundary conditions at $z=0$; some 
possibilities are discussed in 
Refs.~\onlinecite{shen,linder,egger} and \onlinecite{shenoy}. 
However we will follow Ref.~\onlinecite{zhang3}; here the vacuum region is 
also taken to be governed by Eq.~\eqref{ham1}, except that the parameter $m$ 
is large and negative there. We write the Hamiltonian as $H_b = H_0 + H'$, 
where
\bea H_0 &=& - m(z) \tau^z ~-~ i \vz \tau^y \pa_z, \non \\
H' &=& i \vp \tau^x ~(\si^y \pa_x ~-~ \si^x \pa_y), \label{ham2} \eea
where $m(z)$ is a positive (negative) constant for $z<0$ ($z>0$) as stated 
above. Next, we note that $H_0$ has a zero energy eigenvector which is 
localized near the surface; the corresponding wave function is given by
\beq \Psi_1 (x,y,z,t) ~=~ \psi_1 (x,y,t) ~\exp [\frac{1}{\vz} \int_0^z dz' 
m(z')], \label{Psi1} \eeq
where $H_0 \Psi_1 = 0$ implies that the four-component spinor $\psi_1$ must
satisfy
\beq (\tau^z ~+~ i \tau^y) ~\psi_1 ~=~ 0. \label{soln1} \eeq
The identity $(\tau^z - i \tau^y) (\tau^z + i \tau^y) = 2 (I + \tau^x)$
implies that all solutions of Eq.~\eqref{soln1} must satisfy the condition 
$\tau^x \psi_1 = -\psi_1$. 
We now observe that $H'$ commutes with $\tau^x$. The condition 
$\tau^x \psi_1 = -\psi_1$ then implies that the equation $H' \psi_1 = E 
\psi_1$ can be written as $H_{s1} \psi_1 = E \psi_1$, where 
\beq H_{s1} ~=~ -i \vp ~(\si^y \pa_x ~-~ \si^x \pa_y) \label{ham3} \eeq
is the Hamiltonian governing the states at the top surface.
We now see that the solutions of $H_b \Psi_1 = E \Psi_1$ which are localized 
near the top surface must be of the form given in Eq.~\eqref{Psi1}, where
$\psi_1 ~=~ u_1 \exp [i(k_x x + k_y y - Et)]$,
\beq E ~=~ \pm \sqrt{\vp^2 (k_x^2 + k_y^2)}, \label{disp2} \eeq
and $u_1$ satisfies
\bea \vp ~(\si^y k_x ~-~ \si^x k_y) u_1 &=& Eu_1, \label{xx1} \\
\tau^x u_1 &=& - u_1. \label{ham4} \eea
In the basis in which $\si^z$ and $\tau^z$ are diagonal matrices with the
diagonal elements being given by $\si^z = (1,-1,1,-1)$ and $\tau^z = 
(1,1,-1,-1)$, we can show that the normalized solution of the above
equations with $E>0$ takes the form $u_1 = (1/2) (1,(ik_x - k_y)/k,-1,
(-ik_x+k_y)/k)^T$, where $k=\sqrt{k_x^2 + k_y^2}$. We then find that
the expectation value of the spin, ${\vec S} = {\vec \si}/2$, in this
state is given by 
\beq \frac{1}{2} ~\psi_1^\dg {\vec \si} \psi_1 ~=~ \frac{1}{2} ~{\hat z} 
\times {\hat k}. \label{spinmom1} \eeq
This relation between $\la {\vec S} \ra$ and $\hat k$ is called spin-momentum 
locking. 

Next, we can derive an expression for the current $\vec J$ by using the Dirac 
equation $i \pa \psi_1 /\pa t = H_{s1} \psi_1$ and the equation of continuity, 
$\pa \rho /\pa t + {\vec \nabla} \cdot {\vec J} = 0$, where the charge density 
is given by $\rho = \psi_1^\dg \psi_1$. We then find that the state with 
$E>0$ has a current along the $x$ direction given by
\bea {\hat x} \cdot {\vec J} &=& \vp \psi_1^\dg \si^y \psi_1 \label{curr1} \\
&=& \vp ~\frac{\vp k_x}{E}. \label{jx} \eea

Similarly we can consider a side surface lying at $x=0$ such that the system 
(vacuum) lies in the region with $x<0$ ($x>0$). The Hamiltonian in 
Eq.~\eqref{ham1} is taken to be valid in both regions with $m$ being
a positive (large negative) constant respectively. We write the 
Hamiltonian as $H_b = H_0 + H'$, where
\bea H_0 &=& - m(z) \tau^z ~+~ i \vp \si^y \tau^x \pa_x, \non \\
H' &=& ~-~ i (\vz \tau^y \pa_z ~+~ \vp \si^x \tau^x \pa_y). 
\label{ham5} \eea
Now $H_0$ has a zero energy eigenvector localized near the surface,
with a wave function given by
\beq \Psi_2 (x,y,z,t) ~=~ \psi_2 (y,z,t) ~\exp [\frac{1}{\vp} \int_0^x dx' 
m(x')], \label{Psi2} \eeq
where $\psi_2$ is a four-component spinor which satisfies 
\beq (\tau^z ~-~ i \si^y \tau^x) ~\psi_2 ~=~ 0. \label{soln2} \eeq
The identity $(\tau^z + i \si^y \tau^x) (\tau^z - i \si^y \tau^x) = 2 (I +
\si^y \tau^y)$ implies that all solutions of Eq.~\eqref{soln2} must satisfy 
the condition $\si^y \tau^y \psi_2 = -\psi_2$. 
Combining this condition 
with the fact that $H'$ commutes with $\si^y \tau^y$ shows that the equation
$H' \psi_2 = E \psi_2$ can be written as $H_{s2} \psi_2 = E \psi_2$, where 
\beq H_{s2} ~=~ i (\vz \si^y \pa_z - \vp \si^z \tau^z \pa_y), \label{ham6} \eeq
is the Hamiltonian on the side surface. (In deriving Eq.~\eqref{ham6},
we have used the relations $\si^y (\si^y \tau^y) = \tau^y$ and $\si^z \tau^z
(\si^y \tau^y) = - \si^x \tau^x$). Thus, the solutions of $H_b \Psi_2 = 
E \Psi_2$ which are localized near the side surface must be of the form 
given in Eq.~\eqref{Psi2}, where $\psi_2 = u_2 \exp [i(k_y y + k_z z - Et)]$,
\beq E ~=~ \pm \sqrt{\vz^2 k_z^2 + \vp^2 k_y^2}, \label{disp3} \eeq 
and
\bea (-\vz \si^y k_z + \vp \si^z \tau^z k_y) u_2 &=& Eu_2, 
\label{xx2} \\
\si^y \tau^y u_2 &=& -u_2. \label{ham7} \eea
We find that the normalized solution of these equations with $E>0$ has 
an expectation value of the spin given by
\beq \frac{1}{2} ~\psi_2^\dg {\vec \si} \psi_2 ~=~ -~ \frac{\vz k_z}{2E} ~
{\hat y}, \label{spinmom2} \eeq
and a current along the $z$ direction given by
\bea {\hat z} \cdot {\vec J} &=& -\vz \psi_2^\dg \si^y \psi_2 \label{curr2} \\
&=& \vz ~\frac{\vz k_z}{E}. \label{jz} \eea

\subsection{Edge between two surfaces}

We now consider a system which has semi-infinite surfaces given by ($z=0,x<0$) 
and ($x=0,z<0$); these meet along the infinitely long edge $x=z=0$ which runs 
along the $y$ direction. The system lies in the region $x< 0$ and $z<0$. In
this subsection we will ignore the bulk Hamiltonian and work with only the
surface Hamiltonians given in Eqs.~\eqref{ham3} and \eqref{ham6}. This
implicitly assumes that the energy scales of the surface states, such as 
$\vp |\vec k|$ and $\vz |\vec k|$, are both much smaller than the bulk gap. We 
will study two problems, one involving scattering of electrons incident on 
the edge from one of the surfaces, and the other involving the existence of 
states localized along the edges (discussed in the next section). For this 
purpose we first need to find the matching condition at the edge between 
the wave functions on the top and side surfaces.

We saw above that the surface Hamiltonians and conditions have two different 
forms, Eqs.~(\ref{ham3}),(\ref{ham4}) and (\ref{ham6}),(\ref{ham7}), on the 
two surfaces. The current perpendicular to the edge and coming into it 
on the top surface is given by
\beq {\hat x} \cdot {\vec J} ~=~ \vp \psi_1^\dg \si^y \psi_1, \eeq 
while the current going away from the edge on the side surface is given by
\beq - {\hat z} \cdot {\vec J} ~=~ \vz \psi_2^\dg \si^y \psi_2. \eeq
For a plane wave incident on the edge from either surface, the conservation
of current perpendicular to the edge implies that ${\hat x} \cdot 
{\vec J}_1 = - {\hat z} \cdot {\vec J}$, namely,
\beq \vp (\psi_1^\dg \si^y \psi_1)_{x=0} ~=~ \vz (\psi_2^\dg \si^y 
\psi_2)_{z=0}. \label{currcons} \eeq
This implies that the wave functions are related as 
\beq (\psi_2)_{z=0} ~=~ \sqrt{\frac{\vp}{\vz}} ~U (\psi_1)_{x=0}, 
\label{bc} \eeq
where the matrix $U$ must satisfy
\beq U^\dg \si^y U ~=~ \si^y. \label{u1} \eeq
In addition, we require that
\beq U^{-1} \si^y \tau^y U ~=~ \tau^x \label{u2} \eeq
for the consistency of Eqs.~\eqref{ham4} and \eqref{ham7}, 
and 
\beq \si^y U^* \si^y ~=~ U \label{u3} \eeq
in order to maintain time-reversal symmetry.

We find that the most general solution of Eqs.~(\ref{u1}-\ref{u3}) involves 
two real parameters and is given by 
\beq U ~=~ e^{-i(\beta \si^y + \ga \tau^y)} ~e^{i(3\pi/4) \si^y 
\tau^z}. \eeq
(This implies that $U$ is a real and unitary matrix).
However, we can bring the factor of $e^{-i(\beta \si^y + \ga \tau^y)}$ to 
the left hand side of Eq.~\eqref{bc} and use the condition that $\si^y \tau^y
\psi_2 = - \psi_2$ to show that $e^{-i(\beta \si^y + \ga \tau^y)} \psi_2
= e^{-i(\beta - \ga) \si^y} \psi_2$. The general boundary condition 
satisfying Eqs.~(\ref{bc}-\ref{u3}) therefore only has one real parameter 
$\al = \beta - \ga$ and is given by 
\beq (\psi_2)_{z=0} ~=~ \sqrt{\frac{\vp}{\vz}} e^{-i\al \si^y}
e^{i(3\pi/4) \si^y \tau^z} (\psi_1)_{x=0}. \label{umat} \eeq
We note that related matching conditions, but without the parameter $\al$,
have been discussed in Ref.~\onlinecite{brey} and in a TI system with a 
different geometry in Ref.~\onlinecite{kundu}. We will see below that the
parameter $\al$ leads to some non-trivial effects, namely, scattering at
the edge and the appearance of states localized at the edge; such effects
therefore do not appear in Ref.~\onlinecite{brey}.

The parameter $\al$ in Eq.~\eqref{umat} can be given a precise physical 
interpretation as was shown in Ref.~\onlinecite{deb}. Consider a 
$\de$-function potential barrier placed 
along the edge at $x=-\ep$, where $\ep$ is an infinitesimal positive quantity.
If the $\de$-function potential is given by $V_0 \de (x+\ep)$, one can use the
Hamiltonian in Eq.~\eqref{ham3} to integrate through this potential to show 
that the wave function has a discontinuity given by~\cite{deb}
\beq (\psi_1)_{x=0} ~=~ e^{-i\al \si^y} (\psi_1)_{x=-2\ep}, \eeq
where $\al = V_0/\vp$. This gives us an understanding of the parameter $\al$. 
If there is no potential barrier along the edge, we must set $\al = 0$ in 
Eq.~\eqref{umat}. Note that the presence of a $\de$-function potential 
barrier produces a discontinuity in a wave function satisfying the Dirac 
equation, in contrast to the Schr\"odinger equation where the wave function 
remains continuous but its first derivative becomes discontinuous.

We now consider the problem of scattering from the edge. Since there is
translational invariance along the edge, the momentum $k_y$ along that
direction will be conserved. We assume that an electron with energy-momentum 
$(E,k_x,k_y)$, satisfying Eq.~\eqref{disp2}, is incident on the edge from the 
top surface; for definiteness, we assume that $E>0$ and $k_x >0$. The angle of
incidence, $\th_i = \tan^{-1} (k_y/k_x)$, can vary from $-\pi/2$ to $\pi/2$. 
The electron will then be reflected in the top surface with amplitude $r$ 
and energy-momentum $(E,-k_x, k_y)$, and transmitted to the side surface with 
amplitude $t$ and energy-momentum $(E,k_z,k_y)$ satisfying Eq.~\eqref{disp3}.
The conservation of $E$ and $k_y$ imply that $\vz k_z = - \vp k_x$;
the minus sign arises because we require $k_z < 0$ so that the transmitted 
electron moves {\it away} from the edge. If $\psi_i$, $\psi_r$ and $\psi_t$ 
denote the normalized four-component spinors corresponding to the incident, 
reflected and transmitted waves respectively, then they will satisfy the
boundary condition (Eq.~\eqref{umat})-
\beq t \psi_t ~=~ \sqrt{\frac{\vp}{\vz}} e^{-i\al \si^y} e^{i(3\pi/4) \si^y 
\tau^z} (\psi_i ~+~ r \psi_r) \label{psibc} \eeq
at the edge. In general $r$ and $t$ will be functions of $\vec k$, 
$E$ and $\al$. It is clear from Eq.~\eqref{psibc} that changing $\al \to \al 
+ \pi$ only changes the sign of $t$, leaving $|t|^2$ unchanged. Hence $|t|^2$
is a periodic function of $\al$ with period $\pi$, and it is sufficient
to consider $\al$ to lie in the range $[-\pi/2,\pi/2]$. Eq.~\eqref{currcons} 
implies that the reflected and transmitted currents will satisfy current
conservation,
\beq \vp (1 - |r|^2) ~=~ \vz |t|^2. \label{unit} \eeq

In momentum space, the Hamiltonian governing the incident wave on the top 
surface is given by 
\beq H_{s1} ~=~ \vp ~(\si^y k_x ~-~ \si^x k_y), \label{ham8} \eeq
while the Hamiltonian governing the transmitted wave on the side surface is 
given by
\beq H_{s2} ~=~ \vp (\si^y k_x + \si^z \tau^z k_y), \label{ham9} \eeq
where we have used the relation $\vz k_z = - \vp k_x$. We now note that
\beq H_{s2} ~=~ e^{i(3\pi/4) \si^y \tau^z} H_{s1} e^{-i(3\pi/4) \si^y \tau^z}.
\eeq
This implies that if $\al = 0$, then $r=0$ and $t= \sqrt{\vp/\vz}$ will
satisfy both the Dirac equations, $H_{s1} \psi_i = E \psi_i$ and $H_{s2} 
\psi_t = E \psi_t$, and the boundary condition in Eq.~\eqref{psibc}. This 
will be true regardless of the angle of incidence. A similar argument holds 
if the electron is incident from the side surface. We therefore conclude that 
if there is no barrier potential at the edge, there will be no reflection and 
a wave incident from either surface will transmit perfectly to the other 
surface. 

We would like to mention here that our result that there is no reflection
in the absence of a barrier potential differs from the one presented in
Ref.~\onlinecite{deb}. The reason for this difference is that the analysis
in Ref.~\onlinecite{deb} simply assumed certain two-component Dirac equations 
on the two surfaces without considering where these equations come from, 
whereas our present analysis begins from a four-component bulk Hamiltonian 
which appears in a specific system such as $\rm Bi_2 Se_3$. Thus our analysis
shows that the bulk Hamiltonian gives rise to some Dirac equations on the two 
surfaces which are compatible with each other in such a way that there is 
no reflection.

The absence of reflection when there is no barrier potential has an
implication for the conductance across the edge. The conductance can be 
derived as follows. Assuming that the system has a large width in the $y$
direction given by $W$, the current going from the top surface to the side 
surface is given by
\beq I ~=~ e W ~\int \int~ \frac{dk_x dk_y}{(2\pi)^2}~ (-{\hat z} \cdot 
{\vec J}) ~|t|^2, \label{iv1} \eeq
where $e$ is the charge of an electron. Eq.~\eqref{jz} implies that
$-{\hat z} \cdot {\vec J} = -\vz (\vz k_z/E) = \vz (\vp k_x/E) = \vz 
\cos \th_i$ (we recall that $k_z < 0$), where $\th_i$ is the
angle of incidence and $E = \sqrt{\vp^2 (k_x^2 + k_y^2)} = \sqrt{\vz^2 k_z^2 
+ \vp^2 k_y^2}$. We now rewrite $dk_x dk_y = (E/\vp^2) dE d\th_i$. 
The integral over $E$ goes from $\mu_1$ to $\mu_2$ which are the Fermi 
energies (with respect to the Dirac point) in the electron reservoirs 
connected to the top and side surfaces. The voltage applied to a reservoir 
$i$ is related to its Fermi energy as $\mu_i = e V_i$. 
The differential conductance $G=\frac{dI}{dV}$ is the ratio of the change 
in current $dI$ to the change in bias $dV$ when the Fermi energy of TI-1 
is changed from $\mu = eV$ to $\mu + e dV$. This is given by
\beq G ~=~ \frac{e^2 W \mu}{\vp^2 (2\pi \hbar)^2} ~
\int_{-\pi/2}^{\pi/2} ~d \th_i~ \vz \cos \th_i ~|t|^2, \label{iv2} \eeq
where the combination $\mu /(\vp^2 \hbar^2)$ on the right hand side comes from
the density of states of a massless Dirac electron with Fermi energy $\mu$,
and we have restored factors of $\hbar$ here. For the reflectionless case, we 
have $|t|^2 = \vp/\vz$. We then find that $G=G_0$ where
\beq G_0 ~=~ \frac{2e^2 W \mu}{\vp (2\pi \hbar)^2}. \label{conduct} \eeq
The factor of $1/\vp$ in Eq.~\eqref{conduct} can be traced back to the fact 
that we are calculating the conductance across the edge which runs along the 
$y$ direction, and the velocity in that direction is $\vp$. In general, if the 
velocities along the $x$, $y$ and $z$ directions are $v_x$, $v_y$ and $\vz$ 
respectively, the differential conductance would be proportional to $1/v_y$; 
this would come from a product of the density of states for the incident waves
on the top surface, the probability of transmission to the side surface and 
the transmitted current perpendicular to the edge on the side surface which 
are proportional to $E/(v_x v_y)$, $v_x/\vz$ and $\vz$ respectively.

\begin{figure}[htb]
\begin{center} \epsfig{figure=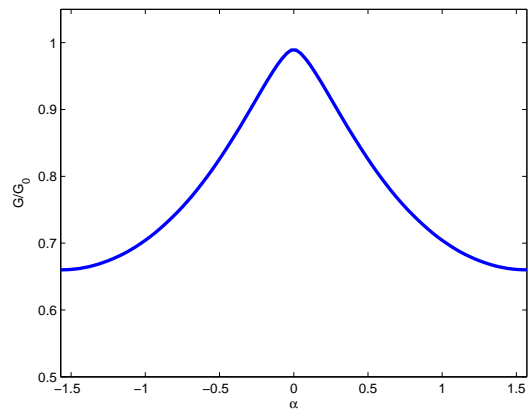,width=8.0cm} \end{center}
\caption{(Color online) Differential conductance $G/G_0$ versus $\al$.}
\label{fig02} \end{figure}

We now study what happens if there is a barrier potential along the edge,
i.e., $\al \ne 0$ in Eq.~\eqref{psibc}. For a wave incident on the top surface
with energy $E>0$ and angle $\th_i$, we find that the transmission amplitude 
is given by
\bea t &=& i ~\sqrt{\frac{\vp}{\vz}} ~\frac{2 \sqrt{2} ~\cos \th_i ~\sqrt{1 - 
\sin \th_i}}{A ~+~ i e^{-i\th_i} B}, \non \\
A &=& (\sin \al + \cos \al) (-i \cos \th_i) \non \\
&& + ~(\sin \al - \cos \al) (1 - \sin \th_i), \non \\
B &=& (\sin \al - \cos \al) (-i \cos \th_i) \non \\
&& - ~(\sin \al + \cos \al) (1 - \sin \th_i). \label{talpha} \eea
Interestingly, $t$ only depends on the velocity ratio $\vp/\vz$ through the 
prefactor $\sqrt{\vp/\vz}$. We can calculate the differential conductance 
using Eqs.~\eqref{iv2} and \eqref{talpha}. In Fig.~\ref{fig02} we show a 
plot of $G/G_0$ as a function $\al$ lying in the range $[-\pi/2,\pi/2]$. It 
has a maximum value of 1 at $\al=0$ and a minimum value of $2/3$ at $\al = 
\pm \pi/2$.

\subsection{Edge states}

In this section, we examine if there are states which are localized along the 
edge. These have a momentum $k_y$ along the $y$ direction, but their wave 
functions decay as $e^{x/\xi_1}$ and $e^{z/\xi_2}$ as we go away from edge on 
the top surface and the side surface respectively. A similar problem has been 
discussed in Ref.~\onlinecite{deb} where only the surface Hamiltonians were 
considered, except that we now have to impose the conditions given in 
Eqs.~\eqref{ham4}, \eqref{ham7} and \eqref{umat}. We find that 
the edge states have an unusual dispersion of the form $E = v |k_y|$, where 
$v$ depends on the barrier parameter $\al$. (We observe that this differs 
from a chiral dispersion where $E/k_y$ has the same sign for both positive 
and negative values of $k_y$). We find that $v = \vp \cos \al >0$ if $-\pi/2 
\le \al < 0$ and $v = - \vp \cos \al <0$ if $0 < \al \le \pi/2$. There are no 
edge states if $\al =0$; thus the absence of edge states goes hand in hand 
with the absence of reflection for the scattering problem. If $\al = \pm 
\pi/2$, the bound states form a flat band with $E=0$. The decay lengths 
$\xi_i$ turn out to be inversely proportional to $|k_y|$; in particular, we 
find that $\xi_1 = 1/|k_y \sin \al|$ and $\xi_2 = (\vz/\vp) \xi_1$. There 
is no edge state at $k_y = 0$ where $\xi_i \to \infty$. Finally, the
expectation value of the spin of the edge states lies in the $x-z$ plane and 
points in a direction which depends on $\al$ and the signs (but not the
magnitudes) of $k_y$ and $E$. For instance, if $k_y, E > 0$, we find that the 
spin points in the direction $-\cos \al ~{\hat x} - \sin \al ~{\hat z}$.

Interestingly, we discover that edge states also appear if a uniform magnetic
field is applied along the $x$ direction and there is no potential barrier
present. A magnetic field adds two terms to 
the Hamiltonian: (i) a Zeeman coupling $-(g \mu_B/2) {\vec \si} \cdot {\vec 
B}$, where $g$ is the gyromagnetic ratio (which we will take equal to 2) and 
$\mu_B = e \hbar/(2m_e c)$ is the Bohr magneton, and (ii) an orbital term 
where we change the momentum operator $-i {\vec \nabla} \to -i {\vec \nabla} 
- (e/c) {\vec A}$, where $\vec A$ is the vector potential. 
For a magnetic field in the $x$ direction, the Zeeman term is given by 
$- (g\mu_B/2) B_x \si^x$. It is convenient to define
\beq b_x ~=~ g \mu_B B_x ~=~ \frac{e\hbar}{m_e c} B_x \eeq
which has the dimensions of energy. We will assume that both $\vp k_y$ and
$b_x$ are small compared to the bulk gap, so that we can use only the 
surface Hamiltonians. Eq.~\eqref{spinmom2} implies that the wave functions
on the side surface have the expectation value $\la \si^x \ra = 0$.
Hence, to first order in $b_x$, we will ignore the term proportional to 
$b_x \si^x$ on the side surface and keep it only on the top surface. 
Turning to the orbital term, we can choose the vector potential to be
${\vec A} = - B_x z {\hat y}$. On the top surface, the vector potential
vanishes since $z=0$; hence we keep the vector potential only on 
the side surface. We will therefore work with the Hamiltonians
\bea H_{s1} &=& -i \vp ~(\si^y \pa_x ~-~ \si^x \pa_y) ~-~ \frac{1}{2} ~b_x 
\si^x, \non \\
H_{s2} &=& i \vz \si^y \pa_z ~+~ \vp \si^z \tau^z (-i \pa_y + \frac{e}{c}
B_x z), \label{hs12} \eea
on the top and side surfaces respectively. We have to impose the conditions
in Eqs.~\eqref{ham4}, \eqref{ham7} and \eqref{umat} (with $\al = 0$). Once 
again we will look for eigenstates with momentum $k_y$ along the $y$ direction 
and energy $E$. We then discover that there are solutions with {\it zero 
energy} with wave functions given by
\bea \psi_1 &=& u_1 e^{i(k_y y - Et) + x/\xi_1}, \label{psi1} \\
\psi_2 &=& u_2 e^{i(k_y y - Et) + z/\xi_2 + sgn (k_y) (e B_x \vp/2 c \vz) 
z^2}, \label{psi2} \eea
on the top and side surfaces (here $sgn$ denotes the signum function), 
provided that \beq b_x ~=~ - ~2 (\lam ~+~ 1) ~\vp k_y, \label{bxky} \eeq
where $\lam$ is positive. Namely, the magnitude of $b_x$ must be larger
than $2\vp |k_y|$, and its sign must be opposite to $k_y$. The decay lengths 
are given by 
\bea \xi_1 ~=~ \frac{2\vp}{|b_x + 2 \vp k_y|} ~~~{\rm and}~~~ \xi_2 ~=~ 
\frac{\vz}{\vp |k_y|}. \label{xi12} \eea 
The forms of the spinors 
$u_1, ~u_2$ in Eqs.~(\ref{psi1}-\ref{psi2}) depend only on the sign of $k_y$, 
and not on the magnitude of $k_y$ or $b_x$. For instance, for $k_y > 0$ and 
$b_x < - 2 \vp |k_y|$, we find that 
\beq u_1 ~=~ \frac{1}{\sqrt 2} ~\left( \begin{array}{c}
0 \\
1 \\
0 \\
-1 \end{array} \right) ~~~{\rm and}~~~
u_2 ~=~ \frac{1}{2} ~\left( \begin{array}{c}
-1 \\
1 \\
-1 \\
-1 \end{array} \right). \label{u12} \eeq
Eqs.~\eqref{u12} show that the $u_1^\dg {\vec \si} u_1 = - {\hat z}$ and
$u_2^\dg {\vec \si} u_2 = 0$. Hence this state will have the expectation
value of $\vec S$ pointing in the $-{\hat z}$ direction. If $k_y < 0$ and 
$b_x > 2 \vp |k_y|$, the edge state will have $\vec S$ pointing in the 
$+{\hat z}$ direction. Note that the edge states that appear at the junction
are mainly due to the Zeeman field. We have included the orbital term only
for the sake of completeness.

If the magnitude of $b_x$ is larger than $2\vp |k_y|$, but its sign is the
same as $k_y$, edge states do not appear along the junction of the top
and side surface but they appear along the junction of other pairs of 
surfaces (such as the bottom surface and the other side surface
shown in Fig.~\ref{fig01}). This can be shown by first deriving the 
Hamiltonians on the other surfaces and then carrying out an analysis 
similar to the one give above.

The zero energy states discussed here occur for reasons which are similar
to the zero energy state which appears in a one-dimensional Dirac equation if 
the mass changes sign at one point~\cite{jackiw}. For a given value of $k_y$, 
we can think of the two Hamiltonians in Eqs.~\eqref{hs12} as being unitarily 
equivalent 
to a one-dimensional Dirac equation with a mass term proportional to $\vp 
k_y + b_x/2$ on one side (the top surface) and to $\vp k_y$ on the other 
side (the side surface). (We are ignoring the orbital term $(e/c) B_x z$ 
here). A zero energy state appears precisely when $\vp k_y + b_x/2$ and 
$\vp k_y$ have opposite signs as we see from Eq.~\eqref{bxky} when $\lam > 0$.

The discussion in this section for 
the various edge states was based entirely on the 
surface Hamiltonians and ignored the bulk Hamiltonian. We would now like to 
study these states more carefully by taking the bulk Hamiltonian into account.
In the continuum, this is a difficult problem to study analytically since the 
edge states will decay into both the surfaces as well as into the bulk. 
Further, the bulk Hamiltonian contains terms of higher order in the momenta 
as we will see in Eq.~\eqref{ham10} below; this too makes an analytical 
calculation difficult. We will therefore study this problem numerically. A 
convenient way of doing this is to consider a lattice version of the model. 
This will be the subject of Secs.~\ref{sec:lattice} and \ref{sec:numerics}.

\section{Lattice model}~\label{sec:lattice}
In this section we will present and numerically study a lattice model for the 
bulk of the system. The lattice model will be used only as a 
way of discretizing the bulk Hamiltonian; the lattice that we will introduce
is {\it not} the same as the original microscopic lattice of $\rm Bi_2 Se_3$. 
A lattice model has several advantages~\cite{imura}: we can obtain 
explicit expressions for the wave functions, both near the surface and in 
the bulk, and we do not have to impose any conditions such as 
Eqs.~\eqref{ham4}, \eqref{ham7} and \eqref{umat}.

In momentum space, a continuum Hamiltonian for $\rm Bi_2 Se_3$ in the bulk 
is given by~\cite{tirev2}
\bea H_b &=& -~ [m - B_1 k_z^2 - B_2 (k_x^2 + k_y^2)] ~\tau^z \non \\
&& +~ \vz \tau^y k_z ~+~ \vp \tau^x (\si^x k_y - \si^y k_x), \label{ham10} \eea
up to second order in $\vec k$ close to the $\Ga$ point of the 
three-dimensional Brillouin zone, where 
$m= 0.28 ~eV$, $B_1 = 6.86 ~eV$\AA$^2$, $B_2 = 44.5 ~eV$\AA$^2$, $\vz = 
2.26 ~eV$\AA, and $\vp = 3.33 ~eV$\AA. Note that if we only keep terms up 
to first order in $\vec k$, Eq.~\eqref{ham10} reduces to Eq.~\eqref{ham1}.
Ref.~\onlinecite{tirev2} gives some additional terms which are proportional 
to the identity matrix and which break particle-hole symmetry; we will ignore 
these terms in the calculations presented here.

We will now introduce a lattice discretization of 
Eq.~\eqref{ham10} as follows.
Since we want to study a system with an infinitely long edge along the $y$ 
direction, we will describe the system using a square lattice in the $x-z$ 
plane and a continuum model in the $y$ direction where the electrons have 
momentum $k_y$. Each site of the square lattice will have four components 
corresponding to Bi and Se with $S^z = \pm 1/2$; the sites will be labeled by 
two integers $(n_x,n_z)$. The lattice spacing will be taken to be 
$a = 9.94$\AA ~which is the height of a single quintuple layer of $\rm Bi_2 
Se_3$~\cite{qi}. We now define a lattice Hamiltonian which
reduces to Eq.~\eqref{ham10} in the limit $k_x a, k_z a \to 0$. This can be 
done by replacing $k_x \to (1/a) \sin (k_x a)$, $k_x^2 \to (2/a^2) [1- \cos 
(k_x a)]$, and similarly for $k_z, k_z^2$. We emphasize that we are using the 
lattice only to discretize the bulk Hamiltonian in Eq.~\eqref{ham10} which is
valid close to zero momentum (the $\Ga$ point). Namely, we will only study 
long wavelength modes using our lattice model. Hence our results can only 
be trusted for those states whose length scales (such as the decay lengths
of the edge states) are much larger than the lattice spacing. In this limit, 
the exact structure of the lattice is not important.

We also want to study the effects of a potential $V_{n_x,n_z}$ placed on 
certain sites of the lattice, and a uniform magnetic field. The magnetic 
field will have both a Zeeman coupling to the electron spin and an orbital 
part. In our numerical studies, we will only consider the effect of the Zeeman
coupling for the following reason. Since the edge states that we will study 
are localized along one of the edges and are therefore quasi-one-dimensional, 
they will not be affected significantly by the orbital term. We also note that
a magnetic field that has only a Zeeman coupling and no orbital coupling can 
be realized in a TI by doping with magnetic impurities~\cite{dopmag}
or by depositing a ferromagnetic layer on the surface~\cite{feroti}.

Putting everything together, the eigenvalue equation on the lattice for 
electrons with momentum $k_y$ takes the form
\bea 
&& - ~(m - \frac{2B_1}{a^2} - \frac{2B_2}{a^2}) \tau^z ~\psi_{n_x,n_z,k_y}
\non \\
&& + ~B_2 k_y^2 \tau^z ~\psi_{n_x,n_z,k_y} \non \\
&& - ~\frac{B_1}{a^2} \tau^z ~(\psi_{n_x,n_z+1,k_y} + \psi_{n_x,n_z-1,k_y})
\non \\
&& - ~\frac{B_2}{a^2} \tau^z ~(\psi_{n_x+1,n_z,k_y} + \psi_{n_x-1,n_z,k_y}) 
\non \\
&& - ~\frac{i\vz}{2a} \tau^y ~(\psi_{n_x,n_z+1,k_y} - \psi_{n_x,n_z-1,k_y})
\non \\ 
&& + ~\frac{i\vp}{2a} \tau^x \si^y ~(\psi_{n_x+1,n_z,k_y} - 
\psi_{n_x-1,n_z,k_y}) \non \\
&& + ~\vp \tau^x \si^x k_y ~\psi_{n_x,n_z,k_y} \non \\
&& + ~(V_{n_x,n_z} ~-~ \frac{g \mu_B}{2} {\vec \si} \cdot {\vec B}) ~
\psi_{n_x,n_z,k_y} \non \\
&& = ~E ~\psi_{n_x,n_z,k_y}, \label{ham11} \eea
where $\psi_{n_x,n_z,k_y}$ denotes the four-component wave function for the 
electron. 
It is interesting to note that the terms involving $B_1$ and $B_2$ help to 
avoid the problem of fermion doubling (i.e., unwanted low-energy modes) at 
$k_x = \pi/a$ and $k_z = \pi/a$ in the bulk spectrum.

It is useful to understand the symmetries of this system. For a given momentum 
$k_y$ in the $y$ direction and energy $E$, the eigenstate $\psi$ will be given 
by $e^{i(k_y y - Et)}$ times $u(n_x,n_z;k_y,E)$ where $u$ is a four-component 
spinor. Assuming that the 
magnetic field $\vec B$ is zero, we find that Eq.~\eqref{ham11} has the 
following symmetries.

\noi (i) Time-reversal symmetry $\cal T$: Eq.~\eqref{ham11} remains invariant
if we complex conjugate all numbers, and transform $k_y \to -k_y$ and 
$u(n_x,n_z,k_y) \to \si^y u^* (n_x,n_z,-k_y)$. It follows that each energy 
eigenvalue $E$ will appear with a double degeneracy corresponding to momenta 
$+k_y$ and $-k_y$. Since $\si^{y*} = - \si^y$, we have ${\cal T}^2 = - I$; 
this implies that there will be a two-fold Kramers degeneracy for $k_y = 0$.


\noi (ii) Complex conjugation symmetry $\cal C$: Eq.~\eqref{ham11} remains 
invariant if we complex conjugate all numbers and transform $u(n_x,n_z,k_y) 
\to u^* (n_x,n_z,k_y)$. This means that the eigenstates can be chosen to
be real if required.



States with $k_y = 0$ deserve a special discussion in the case that
$\vec B =0$. We have already seen that these states have a two-fold Kramers 
degeneracy. Since the Hamiltonian in Eq.~\eqref{ham11} commutes with $\si^y$ 
if $k_y=0$, these states can be 
chosen to be eigenstates of $\si^y$; it then follows that the two states will
have eigenvalues $\pm 1$. To see this, we note that if $\psi_1$ satisfies 
$\si^y \psi_1 = \psi_1$ with eigenvalue $+1$, the other eigenstate will be
$\psi_2 = \si^y \psi_1^*$ and it will satisfy $\si^y \psi_2 = - \psi_2$.

In the next subsection, we will present the numerical results obtained by
solving Eq.~\eqref{ham11}. 
Since we are only using the lattice as a way to study the continuum model,
we can trust our results only if the length scales of variation of the wave 
functions in all three directions are much larger than the lattice spacing $a$.

\section{Numerical results}~\label{sec:numerics}
\begin{figure}[htb]
\begin{center} \epsfig{figure=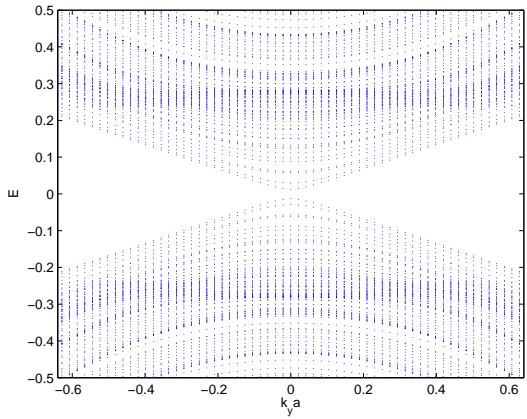,width=8.0cm} \end{center}
\caption{(Color online) $E$ (in $eV$) versus $k_ya$ lying in the range
$[-\pi/5,\pi/5]$ for a $16 \times 16$ lattice, with $V_{n_x,n_z} =0$ at all 
sites.} \label{fig03} \end{figure}

We will begin by considering a $16 \times 16$ lattice with open boundary 
conditions, 
namely, $n_x$ and $n_z$ go from 1 to 16. We first present the results 
obtained when the on-site potential $V_{n_x,n_z} =0$. The energy 
spectrum as a function of the dimensionless parameter $k_y a$ lying in the 
range $[-\pi/5,\pi/5]$ is shown in Fig.~\ref{fig03}. 
[In this section, we are only going to look at states with a limited range of
$k_y a$ close to zero, since it is only for this range that the numerical 
results based on our lattice model can be expected to match with those
obtained using a continuum Hamiltonian]. For each value of 
$k_y$, there are $1024$ energy levels. We see that there is only a small 
energy gap at $k_y =0$ and $E = 0$; this small gap is due to the finite width 
of the system, namely, the fact that the minimum value of 
$k_x$ or $k_z$ is about $\pi/16$ rather than zero. The gap opens up linearly 
as $|k_y|$ increases. Looking at the wave functions, we discover that the 
states closest to $E = 0$ are peaked at the surfaces, namely, near $n_x$ or 
$n_z$ equal to 1 or 16, while those further away from $E=0$ are 
bulk states. More precisely, the states lying within the
range $-m < E < m$ (where $m=0.28 eV$ is half the bulk gap) are surface
states, while the states lying outside this range are bulk states. It is 
visually clear from Fig.~\ref{fig03} that the density of states for 
surface states is less than that for bulk states. This is because, for a given 
value of $k_y$, the surface states are labeled by only one other momentum 
($k_x$ on the top surface and $k_z$ on the side surface), while the bulk 
states are labeled by two other momenta (both $k_x$ and $k_z$); hence the 
number of bulk states is much larger than the number of surface states. 

\begin{figure}[htb]
\begin{center} \epsfig{figure=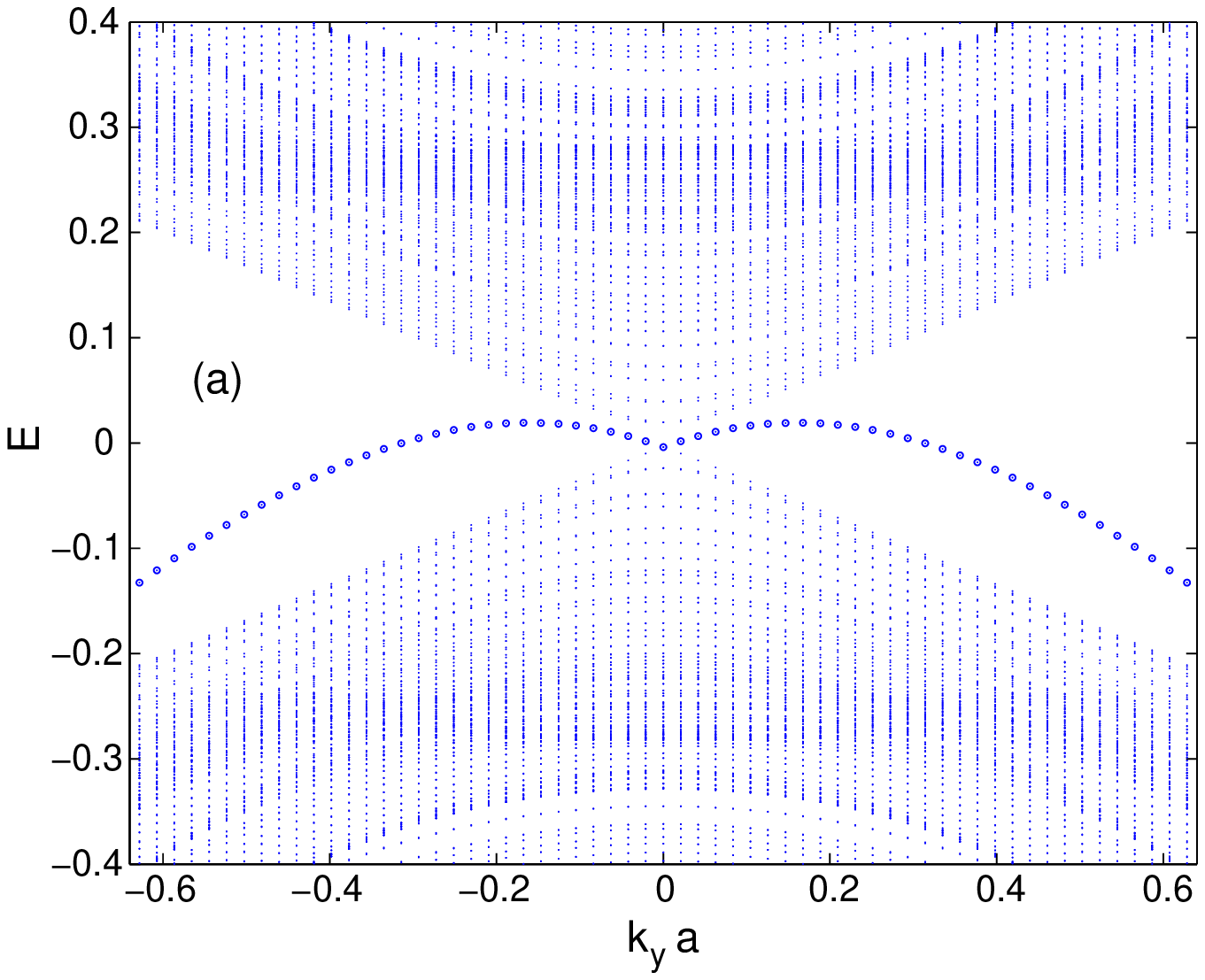,height=4cm,width=8.0cm} \\
\epsfig{figure=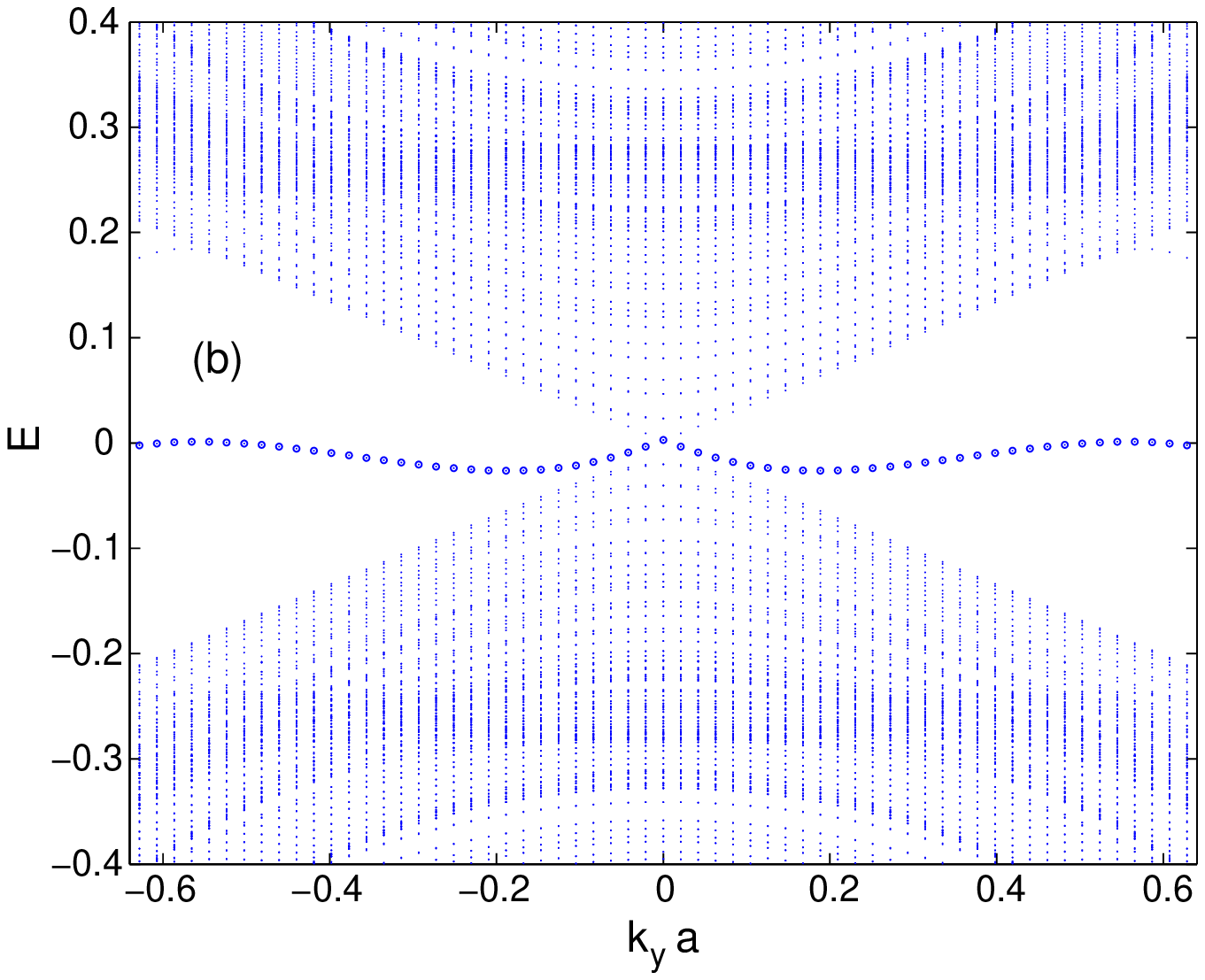,height=4cm,width=8.0cm} \\
\epsfig{figure=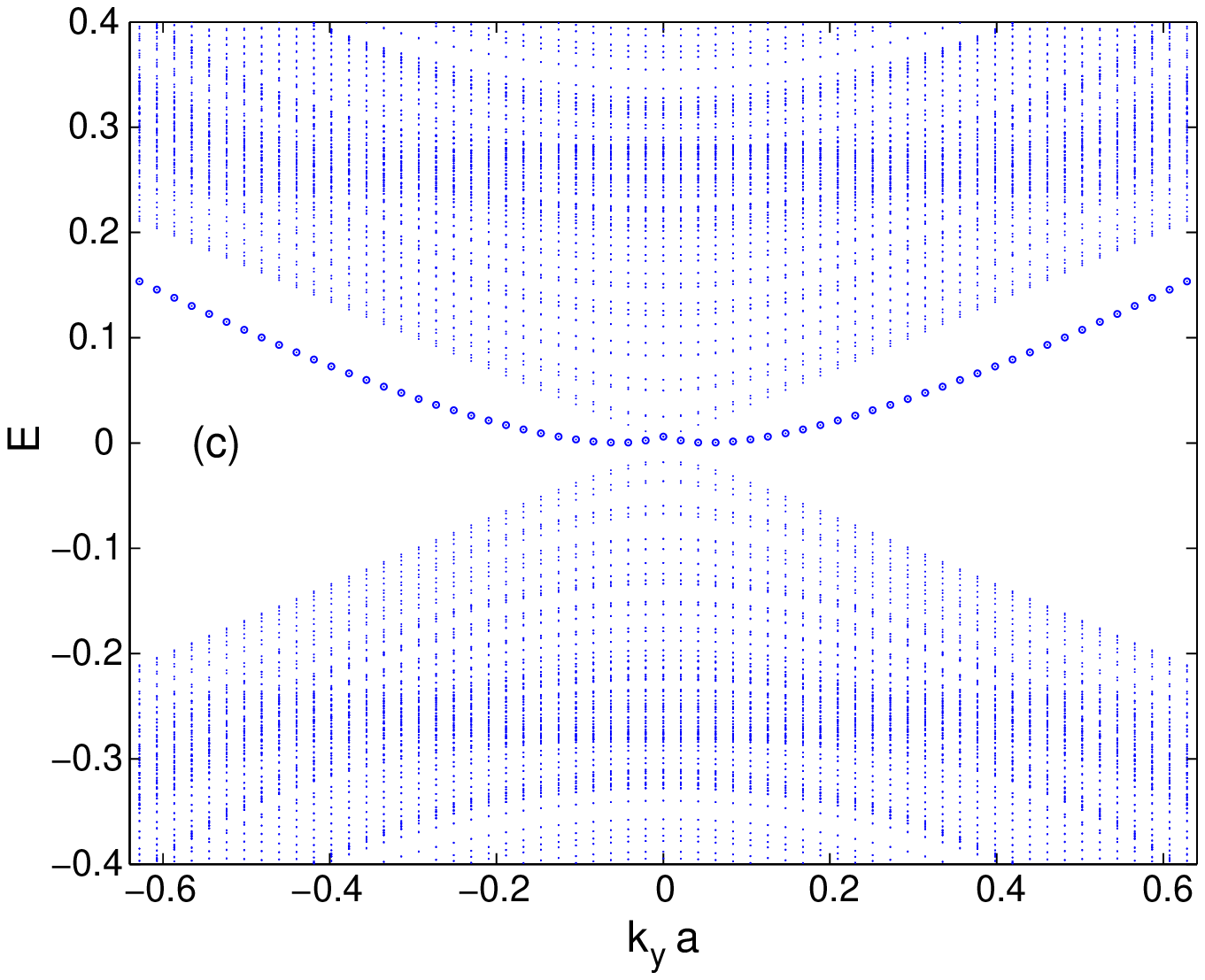,height=4cm,width=8.0cm} 
\end{center}
\caption{(Color online) $E$ (in $eV$) versus $k_ya$ in the range $[-\pi/5,
\pi/5]$ for a $16 \times 16$ lattice, with $V_{16,16}$ being equal to (a) 
$0.5$, (b) 1 and (c) $1.5 ~eV$.} \label{fig04} \end{figure}

We now introduce a potential at one corner of the lattice. We have
considered three cases corresponding to $V_{n_x,n_z} =0.5$, 1 and $1.5~ eV$
at $(n_x,n_z)=(16,16)$ and zero everywhere else. 
The energy spectra as functions of $k_y a$ are shown in Figs.~\ref{fig04} 
(a-c). In each figure we see a set of states (shown by thick blue dots) which 
stay close to $E = 0$ for $k_y a$ lying in the range $[-\pi/5,\pi/5]$. 
We find that these states are non-degenerate, except at $k_y = 0$ where there 
is a Kramers degeneracy. The wave functions of these states are localized at 
the corner of the lattice where the potential is present; hence we will refer 
to these as corner states in this section. We note that $n_z = 16$ ($n_x = 16$)
correspond to the top (side) surface discussed in Sec.~\ref{sec:continuum} A; 
hence the corner $(n_x,n_z)=(16,16)$ corresponds to the edge between those 
two surfaces discussed in Sec.~\ref{sec:continuum} C. 

The dispersion found numerically agrees qualitatively (but not quantitatively)
with the one found analytically in Sec.~\ref{sec:continuum} C where it was 
shown that the energy is of the form $E = v |k_y|$, where $v$ changes from 
negative to positive values as the potential parameterized by $\al$ is 
increased. Namely, we see from Figs.~\ref{fig04} (a-c) that the energy is the 
same for $k_y$ and $-k_y$, and that it curves downwards for $V=0.5 ~eV$, is 
quite flat for $V=1 ~eV$ and curves upwards for $V=1.5 ~eV$. Thus the 
numerical calculation on the lattice confirms the result obtained analytically
that the parameter $\al$ describes a potential at the edge between two 
surfaces and leads to the presence of edge states whose dispersion changes 
with $\al$. We should remark here that the numerical results are not 
expected to quantitatively match
the analytical results for very small values of $|k_ya|$ where the decay 
length of the corner states becomes of the order of the system size (namely,
these cease to be corner states). Further, the fact that the numerically
obtained dispersion is non-linear rather than linear in $k_y$ may be due
to the terms proportional to $B_1$ and $B_2$ in Eq.~\eqref{ham10}.

\begin{figure}[htb]
\begin{center} \epsfig{figure=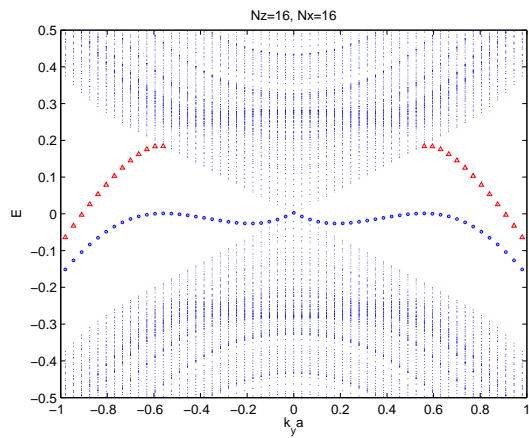,width=8.0cm} \end{center}
\caption{(Color online) $E$ (in $eV$) versus $k_ya$ in the range $[-1,1]$
for a $16 \times 16$ lattice, with $V_{n_x,n_z} =1 ~eV$ at $(n_x,n_z)=
(16,16)$. Two sets of corner states are shown by large blue dots (these states
exist close to $k_y=0)$ and red triangles (these only exist for $|k_y a| 
\gtrsim 0.6$).} \label{fig05} \end{figure}

We note that any potential on a lattice necessarily has a finite width of at 
least one lattice spacing, and is therefore not identical to a $\de$-function 
potential. Hence a lattice model with a potential will sometimes show 
additional sets of corner states which lie away from $k_y a =0$. This is in 
contrast to the continuum model with a $\de$-function barrier where we get 
only one set of corner states whose momenta come arbitrarily close to $k_y = 
0$ where the energy also tends to zero. (We recall that states obtained in a 
lattice calculation have a counterpart in the continuum only in the limit 
$k_y a \to 0$). An example of such additional states is shown by the two 
sets of red triangles in Fig.~\ref{fig05}. In this figure the 
parameters are the same as in Fig.~\ref{fig04} (b) but the range of $k_y a$
has been increased to $[-1,1]$ in order to show these extra states.
We also note that the states shown by large blue dots which approach $E=0$ as
$k_y \to 0$ go quite far away from zero energy as $k_y a$ becomes large; this 
is due to the non-linear terms proportional to $B_1$ and $B_2$ in 
Eq.~\eqref{ham10}. In the rest of our discussion we will only concentrate on 
the corner states which lie close to zero momentum and zero energy since these
are the direct counterparts of the edge states found in the continuum model.


\begin{figure}[htb]
\begin{center} \epsfig{figure=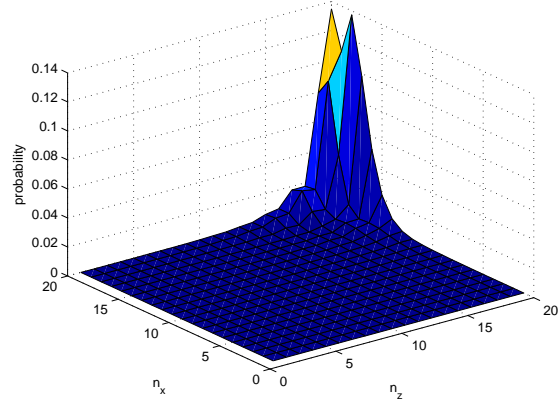,width=8.0cm} \end{center}
\caption{(Color online) Surface plot of the probability density versus $n_x$
and $n_z$ for the corner state with $k_y a =0.2$, for a $20 \times 20$ lattice
with $V_{20,20} =1 ~eV$.} \label{fig06} \end{figure}

A surface plot of the probability for the wave function of the corner state 
with $k_y a = 0.2$ is shown in Fig.~\ref{fig06}. We consider each 
lattice site labeled as $(n_x,n_z)$ and calculate the probability density
namely, $P_{n_x,n_z,k_y} = \psi_{n_x,n_z,k_y}^\dg \psi_{n_x,n_z, k_y}$. This 
is then shown as a surface plot versus $n_x$ and $n_z$.

\begin{figure}[htb]
\begin{center} \epsfig{figure=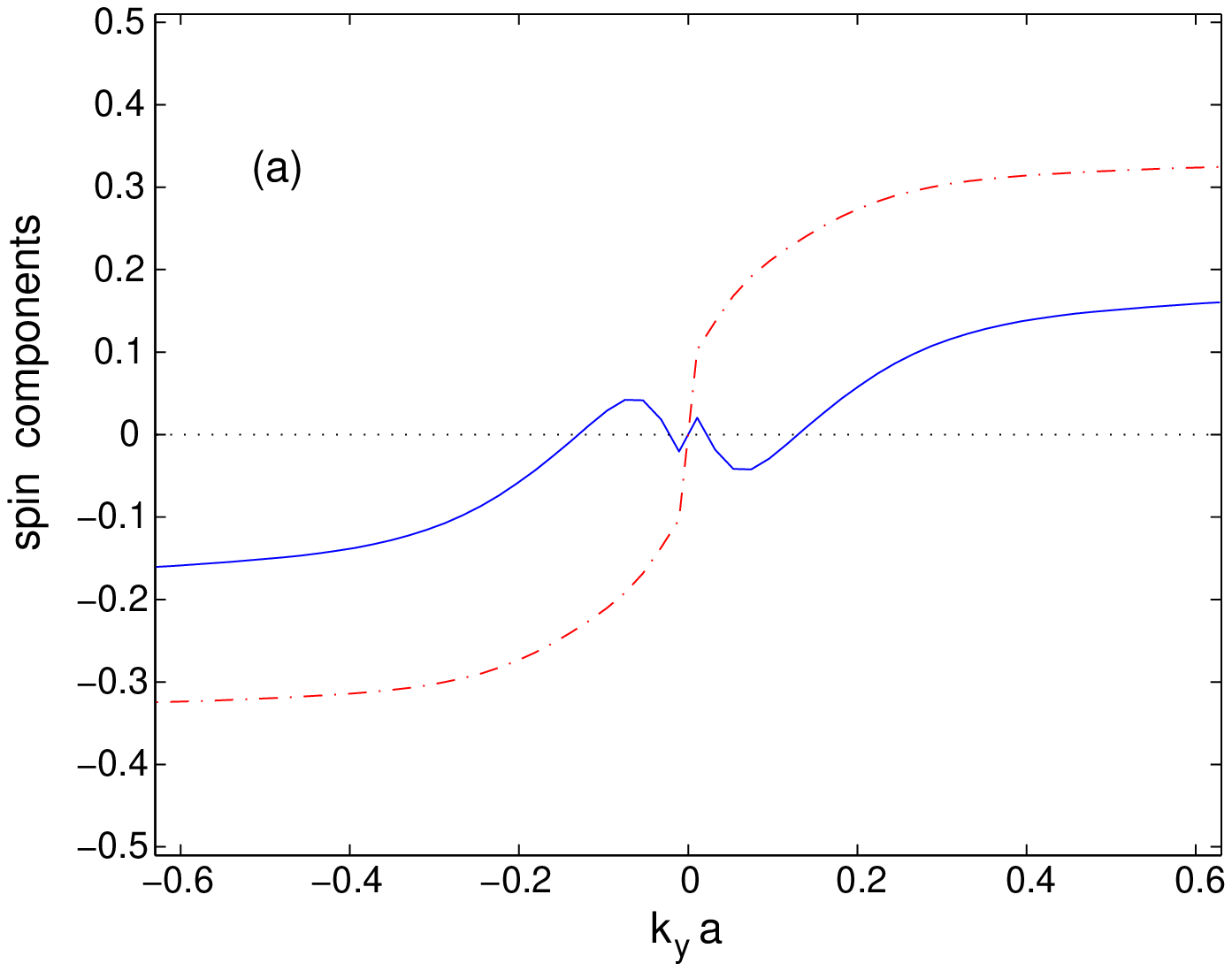,height=5cm,width=8.0cm} \\
\epsfig{figure=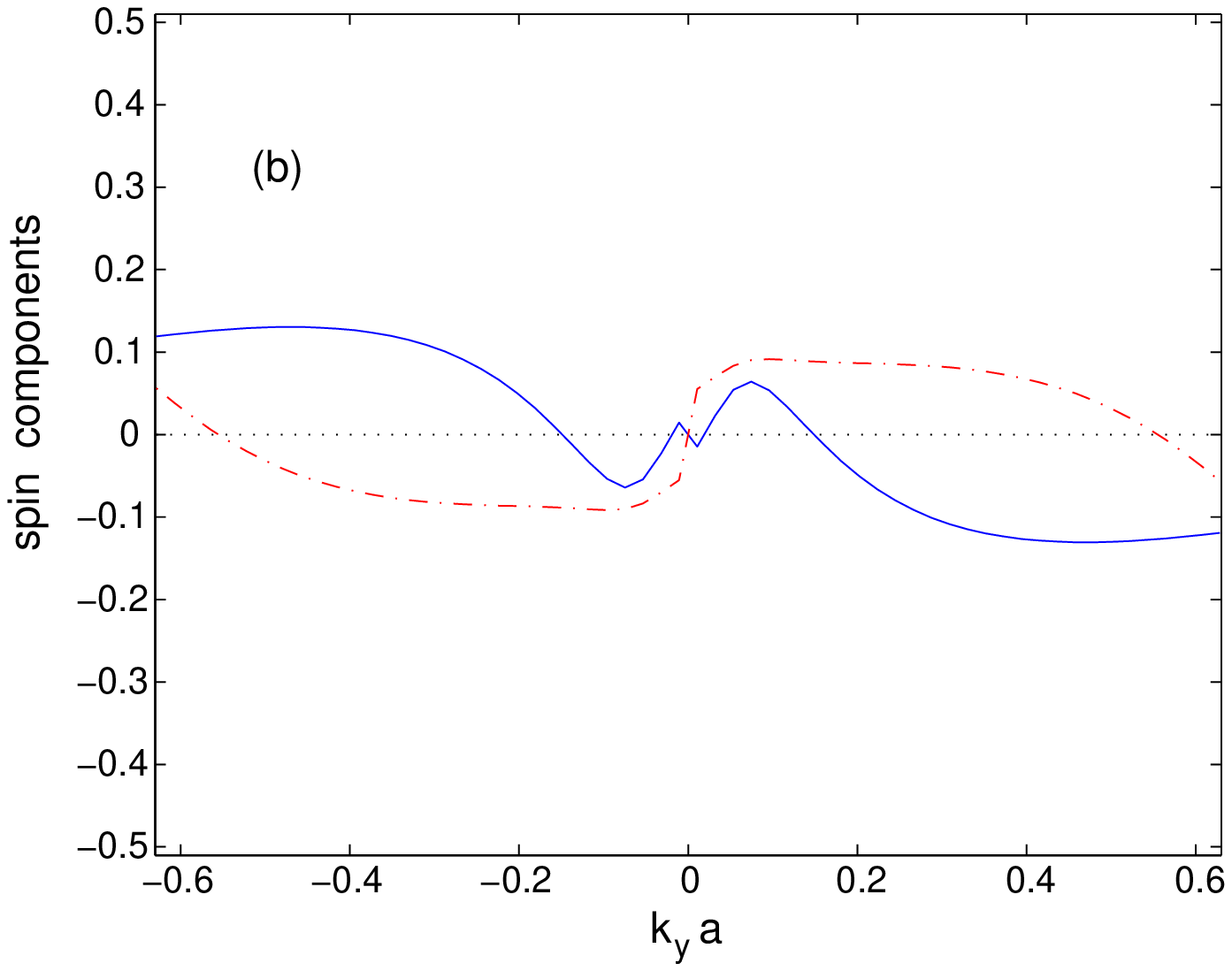,height=5cm,width=8.0cm} \\
\epsfig{figure=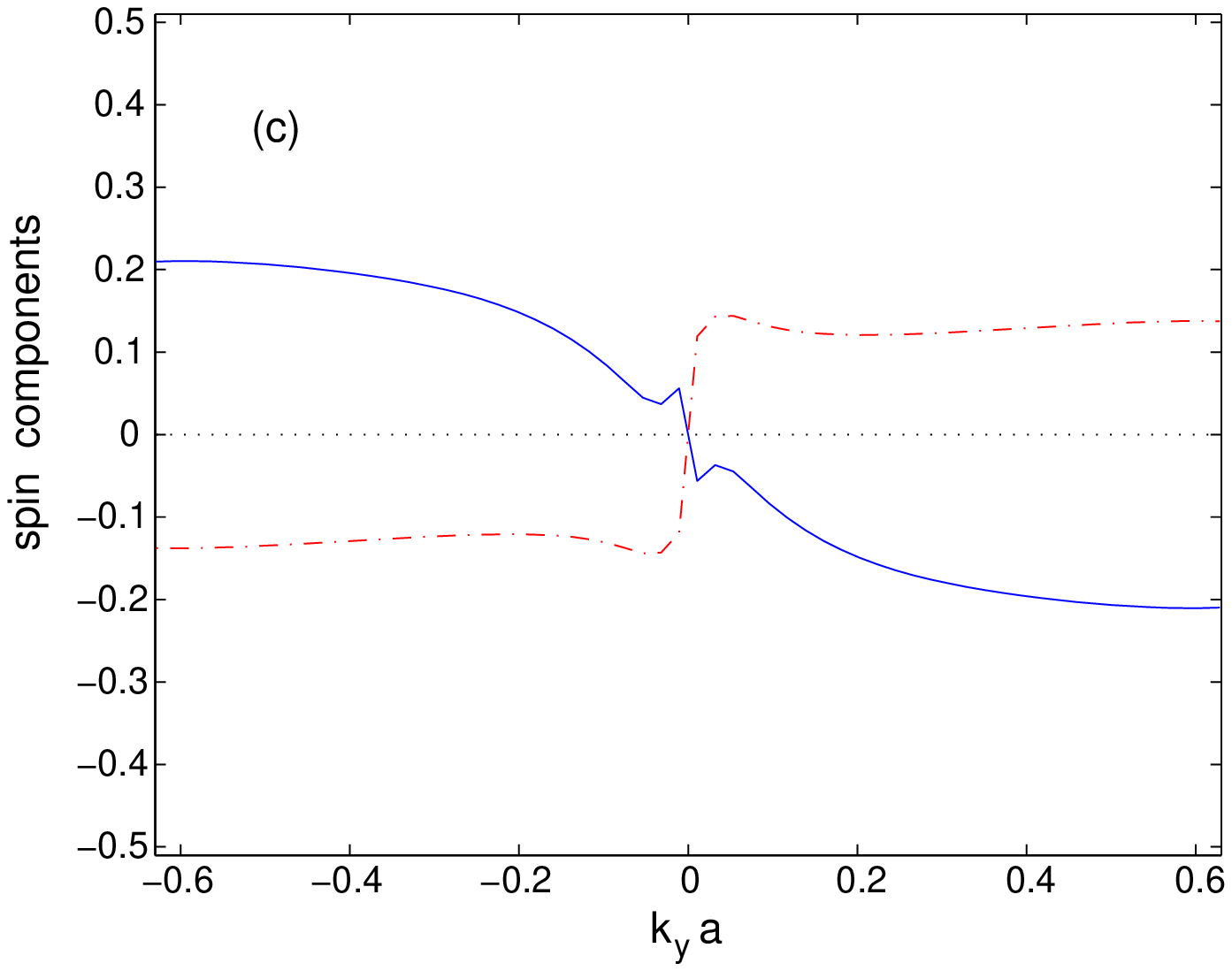,height=5cm,width=8.0cm} 
\end{center}
\caption{(Color online) Spin components $S^x$ (blue solid), $S^y$ (black 
dotted) and $S^z$ (red dash dot) versus $k_ya$ for the corner states of a 
$20 \times 20$ lattice, with $V_{20,20}$ being equal to (a) $0.5$, (b) 1 
and (c) $1.5 ~eV$. } \label{fig07} \end{figure}

It is interesting to look at the spin structure of the corner state, namely, 
the expectation values of $S^a = \si^a/2$. Fig.~\ref{fig07} shows the 
expectation values $\la S^x \ra$, $\la S^y \ra$ and $\la S^z \ra$ as 
functions of $k_ya$ for the corner state of a $20 \times 20$ lattice with 
$V_{20,20} =0.5$, 1 and $1.5$. (The rapid variations of some of the 
quantities near $k_y = 0$ may not be significant since, as remarked earlier, 
the decay length of the corner state becomes of the order of the system 
size when $k_y \to 0$. The properties of the corner states can no longer 
be found reliably when that happens). The figures shows that $\la S^x \ra$ 
and $\la S^z \ra$ flip sign under $k_y \to - k_y$, while $\la S^y \ra = 0$ 
for all values of $k_y$. (An exception to this occurs at $k_y = 0$ where there
are two states which are eigenstates of $S^y$ with eigenvalues $\pm 1/2$). 
These statements can be proved as follows.
The time-reversal symmetry $u (n_x,n_z,k_y) \to \si^y u^* (n_x,n_z,-k_y)$ 
imply that $\la S^x \ra$ and $\la S^z \ra$ change sign under $k_y \to - k_y$ 
because $\si^x$ and $\si^z$ anticommute with $\si^y$. The complex conjugation 
symmetry $\cal C$ discussed in Sec.~\ref{sec:lattice} implies that $\la S^y 
\ra = 0$ for each value of $k_y$ because $\si^{y*} = - \si^y$. 

Fig.~\ref{fig07} implies that the nature of the spin-momentum locking 
of the corner states differs from that of the surface states. The direction 
of spin lies in the $x-z$ plane and depends on both the sign and magnitude of
$k_y$ and the barrier potential $V$. However the property that the sign of 
$\la {\vec S} \ra$ flips under $k_y \to - k_y$ holds for both surface 
states and corner states.

\begin{figure}[htb]
\begin{center} \epsfig{figure=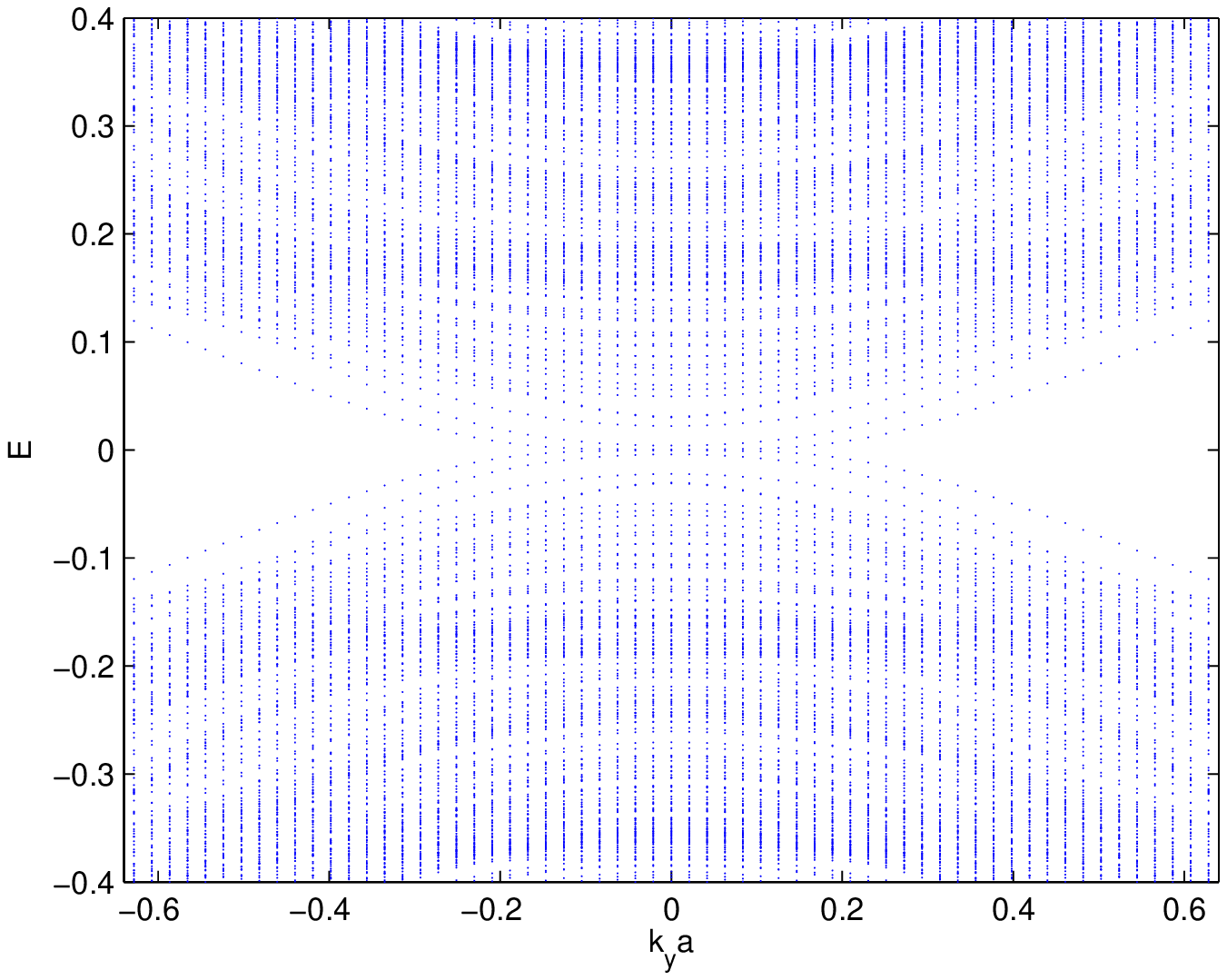,width=8.0cm}
\end{center}
\caption{(Color online) $E$ (in $eV$) versus $k_ya$ for a $16 \times 16$ 
lattice with $g\mu_B B_x = -0.2$. The corner states have an energy close to
zero in the range $|k_y a | \lesssim 0.15$.} \label{fig08} \end{figure}

\begin{figure}[htb]
\begin{center} \epsfig{figure=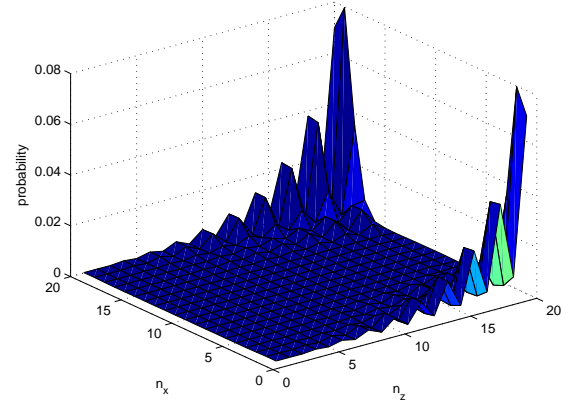,width=8.0cm} \end{center}
\caption{(Color online) Surface plot of the probability versus $n_x$ and
$n_z$ for states at the corners $(0,20)$ and $(20,20)$ with $k_y a =0.1$, for 
a $20 \times 20$ lattice with $g\mu_B B_x = -0.2$.} \label{fig09} \end{figure}

Next, we find that a magnetic field along the $x$ direction can also produce 
corner states with finite momentum. The energy-momentum dispersion is shown 
in Fig.~\ref{fig08}. The energy of these states goes to zero in the 
thermodynamic limit as we will discuss below.
For each value of $k_y$ and $B_x$ consistent with Eq.~\eqref{bxky}, there 
are two states each of which lies at two out of the four corners. [There are 
certain linear combinations of these two states which lie at only one of the 
corners. Due to the finite size of our systems, there is tunneling between 
the two corners; hence the energies of the two states split to give some 
non-zero values $\pm E$ instead of lying exactly at zero. The splitting goes 
to zero exponentially as the system size is increased]. Fig.~\ref{fig09} shows
the probability density for states appearing at the corners $(n_x,n_z)=(0,20)$
and $(20,20)$ for $g\mu_B B_x = -0.2$ and $k_y a = 0.1$.
Note that the decay lengths of these states along the $x$ and $z$ directions
are quite different from each other, as suggested by Eq.~\eqref{xi12}.

\begin{figure}[htb]
\begin{center} \epsfig{figure=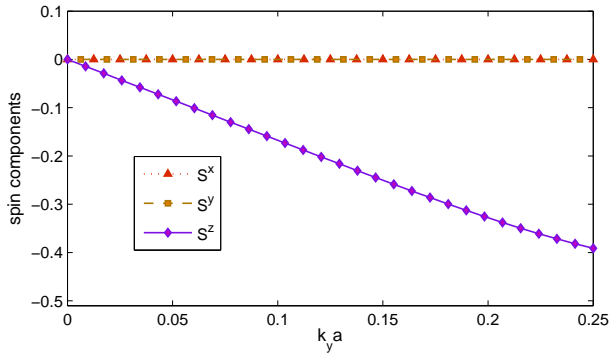,width=9.0cm}
\end{center}
\caption{(Color online) Spin components $S^x$, $S^y$ and $S^z$ versus $k_ya$ 
in the range $[0,0.25]$ for the states at the corner $(20,20)$ of a $20 
\times 20$ lattice with $g\mu_B B_x = -0.2$. The lines for $\la S^x \ra = 
\la S^y \ra = 0$ coincide.} \label{fig10} \end{figure}

Fig.~\ref{fig10} shows the expectation values $\la S^x \ra$, $\la S^y \ra$ 
and $\la S^z \ra$ as functions of $k_ya$ for the states localized at the 
corner $(n_x,n_z)=(20,20)$ of a $20 \times 20$ lattice with $b_x \equiv 
g\mu_B B_x = -0.2$. We have numerically constructed such states by taking 
linear combinations of two energy eigenstates each of which has states at 
two corners as shown in Fig.~\ref{fig09}. 
We have chosen $k_y > 0$ and $b_x < 0$ so that the states appears near the 
corner at $(20,20)$ where the analysis around Eqs.~(\ref{hs12}-\ref{u12}) is
applicable. 
In accordance with the statement made after Eq.~\eqref{u12} for $k_y > 0$ and
$b_x < - 2 v_p |k_y|$, we see that $\la S^x \ra = \la S^y \ra = 0$ while 
$\la S^z \ra < 0$. Interestingly we find that $\la S^z \ra$ varies 
almost linearly with $k_y$. If we choose $k_y < 0$ and $b_x < 0$ with 
$b_x < - 2 v_p |k_y|$, we find states appearing at the other two corners
$(0,0)$ and $(20,0)$. To summarize, we see that a magnetic field in the 
$x$ direction can produce corner states with a momentum in the $y$ 
direction and a spin in the $z$ direction.

\section{Summary and discussion}~\label{sec:summary}

In this paper, we have studied various properties of the surface and edge
states of a three-dimensional TI, both analytically using a continuum theory
and numerically using a lattice model. Analytically, we first found the 
Hamiltonians which 
govern the states on different surfaces (in particular, on the top surface 
and a side surface) of the system starting from a continuum bulk Hamiltonian 
which involves four bands. We then found the most general boundary condition 
that the wave function has to satisfy at the junction of the top and a side 
surface; this turns out to involve a single parameter $\al$ which can be 
interpreted as arising from a barrier running along the junction. Using the 
surface Hamiltonians and the boundary condition, we obtained the following
results.

\noi (i) We found the dependence of the differential conductance $G$
through the junction on the barrier parameter $\al$. For $\al=0$, the
Hamiltonians on the top and side surfaces are related by the same unitary
transformation as the one relating the wave functions. Using this, we have 
shown that there is no reflection from the junction and the transmission is 
perfect; hence $G$ has the maximum possible value at $\al = 0$. The minimum 
value of $G$ occurs at $\al = \pm \pi/2$. 

\noi (ii) We have shown that the barrier gives rise to edge states which 
propagate as plane waves along the junction and decay exponentially away
from the junction. The energy, decay length and spin of these states are
related to the momentum $k_y$ along the junction.
In particular, the energy is given by $v|k_y|$, where $v$ depends on the 
barrier parameter $\al$. For the special values of the barrier parameter 
$\al =\pm \pi/2$, the energy is zero for all $k_y$. For any value 
of $\al$, the energy of the edge states is less than that of the surface 
states with the same momentum $k_y$. This implies that momentum conserving
scattering cannot cause a transition between an edge state and a surface 
state. However momentum non-conserving scattering (which may arise from
impurities lying on or near the barrier) can cause such transitions.

\noi (iii) We have shown that a magnetic field applied in a direction 
perpendicular to the junction of the top and a side surface can also give rise
to localized edge states if a dimensionless ratio between the field $B_x$ and 
the edge momentum $k_y$ is larger than some value. These edge states have zero
energy and a spin which points in the $\pm z$ direction, i.e., perpendicular 
to both the magnetic field and the momentum along the junction.

To understand better the properties of the edge states found analytically
from the continuum theory, we returned to the 
bulk Hamiltonian and studied it numerically as follows. We introduced a 
lattice in the $x-z$ plane keeping the $y$ direction to be a continuum with 
translational invariance; this enables us to maintain $k_y$ as a good quantum
number. We used this model to numerically study different kinds of 
edge states. The advantages of a lattice model are that we work directly
with the bulk Hamiltonian which is valid everywhere (unlike the surface
Hamiltonians which are only valid when we are on a particular surface and 
are far from a junction of two surfaces), and we do not have to impose any
unusual boundary conditions at a surface or a junction of two surfaces.
Our numerical calculations gave the following results.

\noi (iv) In the absence of any on-site potential, the energy spectrum shows 
two kinds of states. The states lying within the bulk gap, i.e., in the range 
$-m<E<m$ where $m=0.28 ~eV$, are peaked at the surfaces and hence are surface 
states. The states lying outside this range are bulk states. The spectrum 
shows a small gap at $k_y=0$ due to the finite size of the system.

\noi (v) In the presence of a potential at one corner of the $x-z$ lattice
(i.e., at one edge of the three-dimensional system), a new 
set of states appears whose wave functions are localized at that corner. 
We refer to these as corner states for convenience; they correspond precisely 
to the edge states found using the continuum surface Hamiltonians as 
described above. The dispersion for these states obtained numerically 
for the bulk lattice system agrees qualitatively with the dispersion
derived using the continuum theory for the surface states. 
The expectation value of the spin of these states is found to lie in the 
$x-z$ plane; the $y$ component of the spin is zero which can also be proved 
by a symmetry argument. Thus the spin is perpendicular to the momentum,
although the direction of the spin of these states is different from that
of the surface states on the adjacent surfaces. 

\noi (vi) We studied the effect of a uniform magnetic field applied along 
the $x$ direction,
keeping only the Zeeman coupling in the lattice Hamiltonian. We find corner 
states under similar conditions on the relative strengths of the magnetic 
field and the momentum as found in the continuum model. For the choice $k_y 
> 0$ and $g\mu_B B_x < 0$, we found numerically that the expectation value of 
spin has a non-zero component only in the $-z$ direction, and its magnitude 
varies linearly with $k_y$.

We have numerically checked that edge (corner) states also occur in two other 
three-dimensional TIs, $\rm Bi_2 Te_3$ and $\rm Sb_2 Te_3$. For 
this purpose, we have used the values of the parameters in Eq.~\eqref{ham10} 
given in Ref.~\onlinecite{tirev2} for these two materials. There are some
differences between the energy spectra in the three materials. 
While the surface and bulk states are separated from each other in energy for 
$\rm Bi_2 Se_3$ and $\rm Sb_2 Te_3$, they occur at the same energies for 
$\rm Bi_2 Te_3$. (This is in agreement with known results; see Figs.~20 (b-d) 
of Ref.~\onlinecite{tirev2}). Similarly, the values of the barrier potential 
where the edge state dispersion is almost flat and the ranges of the magnetic 
field value for which edge states appear are different in the three materials. 
However, the qualitative observation that barrier potentials at junctions of 
two surfaces or a magnetic field applied in the $x$ direction can give rise 
to edge states holds in all these materials.

Our results can be experimentally tested as follows. First of all, a barrier 
potential can be applied near a junction of two surfaces by placing a gate 
close to the junction and tuning the gate voltage. Then spin-resolved ARPES 
can be used to find the energy dispersion and spins of the different edge 
states. This method cannot be easily used in the presence of a magnetic 
field since the field would affect the trajectories of the electrons emitted 
from the surface. A second method would be to measure the local density of 
states using the tunneling conductance from a spin-polarized STM tip 
placed very close to the junction. If the local density of 
states is found to be higher when a potential is applied to the junction (or
if a magnetic field is applied) compared to the case of no potential and
no magnetic field, this would provide evidence for the edge states.
Finally, we can measure the differential conductance between two point 
contacts placed at the two end points of the junction (i.e., the conductance 
in the $y$ direction); a significant value of this conductance and its 
variation with the barrier potential or a 
magnetic field would provide indirect evidence for edge states. Note that 
since the edge states carry a spin which is opposite for opposite edge 
momenta $+k_y$ and $-k_y$, a non-zero charge conductance along the junction 
also implies a non-zero spin conductance.

In our numerical calculations we have considered a lattice in the $x-z$ plane 
and continuum along the $y$ direction. It would be interesting to study the 
effects of impurities present at the junction 
on the edge states. Since the presence of impurities breaks 
the translational invariance along the $y$ direction, $k_y$ would no longer 
be a good quantum number. Hence it would be necessary to use a lattice model 
in all three directions. This may be numerically quite challenging.

It would also be interesting to use the lattice model to numerically study the
effects of the orbital coupling to a magnetic field. On a lattice, such a 
coupling can be introduced through the phase in the couplings between nearest 
neighbors following the Peierls prescription. This can be used to study 
Aharonov-Bohm oscillations due to states which go around all the four surfaces
in Fig.~\ref{fig01}, as was studied experimentally in Ref.~\onlinecite{peng}.

\acknowledgments
We thank Sourin Das, Taylor Hughes, Nitin Samarth and Vijay Shenoy for 
discussions. D.S. thanks DST, India for support under Grant No. 
SR/S2/JCB-44/2010.

\end{document}